\documentclass[twocolumn,prl,aps,floatfix,showpacs]{revtex4}
\usepackage{graphicx}
\usepackage{epsfig}
\usepackage{verbatim}
\usepackage{textcomp}
\usepackage{color}
\usepackage{colordvi}
\newcommand {\hide}[1]{\null}

\newcommand {\kb}{k_{\rm B}}

\newcommand{\MCconc}{3:1}
\newcommand{\Saltconc}{1:1}
\usepackage[T1]{fontenc}

\begin{document}
\title{Effect of Mono- and Multivalent Salts on Angle-dependent
Attractions \linebreak
between Charged Rods}
\author{Kun-Chun Lee$^1$,
Itamar Borukhov$^1$,
William M.\ Gelbart$^1$,
Andrea J.\ Liu$^1$,
and
Mark J.\ Stevens$^2$
}

\affiliation{$^1$Dept. of Chemistry and Biochemistry, University
of California, Los Angeles, CA 90095-1569 \\
$^2$Sandia National Laboratories, Albuquerque, NM 87185-1411}


\begin{abstract}
Using molecular dynamics simulations we examine the effective
interactions
between two like-charged rods as a function of angle and separation.
In particular, we determine how the competing
electrostatic repulsions and
multivalent-ion-induced attractions
depend upon concentrations
of simple and multivalent salt.  We find that
with increasing multivalent salt
the stable
configuration of two rods
evolves from isolated rods to aggregated perpendicular rods
to aggregated parallel rods;
at sufficiently high concentration, additional
multivalent salt reduces the attraction.   Monovalent salt enhances the
attraction near the onset of aggregation,
and reduces it at higher concentration of multivalent salt.

\end{abstract}


\maketitle


Multivalent-ion-induced aggregation of stiff polyelectrolytes has
been studied extensively in
recent years for at
least two important reasons: it arises from correlations not
included at the mean-field (Poisson-Boltzmann) level
\cite{OOSAWA:68,GULDBRAND.NILSSON.EA:86,
ROUZINA.BLOOMFIELD:96*1,
GRNBECH-JENSEN.MASHL.EA:97,HA.LIU:97,
ARENZON.STILCK.EA:99,SOLIS.DE-LA-CRUZ:99,
STEVENS:99,NGUYEN.ROUZINA.EA:00},
and it lies at the heart of
biological phenomena such as cytoskeleton
re-organization \cite{POLLARD:86,JANMEY.HVIDT.EA:90,SACKMANN:96} and
DNA packaging
\cite{WIDOM.BALDWIN:83,BLOOMFIELD:91}.
Most studies have focused exclusively
on the interaction between
charged rods that are parallel.  A few theoretical calculations have
allowed for non-parallel orientations \cite{HA.LIU:99,STILCK.LEVIN.EA:02},
but only at a single multivalent salt concentration.  Recent work
\cite{BORUKHOV.BRUINSMA:01,BORUKHOV.BRUINSMA.EA:03,WONG.LIN.EA:03}, however,
suggests that the
{\it evolution} of angle-dependent attractions with changing multivalent salt
concentration is crucial to phase behavior.
For example, at intermediate multivalent salt concentrations
F-actin solutions can form lamellar phases of
stacked rafts where each raft consists of two layers of mutually
perpendicular actin filaments,
but at higher
multivalent salt concentrations they form bundles of nearly
parallel filaments \cite{WONG.LIN.EA:03}.  Similar physics
may apply to transitions from networks to bundles of F-actin in
the cytoskeleton
\cite{BORUKHOV.BRUINSMA.EA:03}.


In this study
we use explicit-ion, continuum-dielectric molecular dynamics (MD)
simulations to examine how the
concentrations of monovalent and multivalent salts
affect the angle- and distance-dependent effective
potential between two charged rods.  A threshold
concentration of multivalent salt is needed for the two rods
to attract.
We find that the preferred configuration is perpendicular just above the
threshold, and parallel at higher concentrations of multivalent salt,
in agreement with the experiments on F-actin solutions cited above
\cite{WONG.LIN.EA:03}.
Furthermore, monovalent salt can lower the concentration of
multivalent salt needed for attraction, suggesting that monovalent
salt may induce aggregation.  Finally, we find that the rods can still
attract even when they are overcharged \cite{NGUYEN.ROUZINA.EA:00} (i.e. the
sign of their
effective charge is reversed by condensed counterions) at high multivalent
salt concentrations, and that the attraction weakens with
increasing overcharging.  However, we never observe strong overcharging
because above a certain concentration of multivalent salt the added
multivalent ions simply form complexes with monovalent co-ions
in solution \cite{GONZALEZ-MOZUELOS.OLVERA-DE-LA-CRUZ:03}.

In our simulations each rod is composed of 64 spherical
monomers, each carrying a charge of -1 in units of the electronic
charge $e$, separated at fixed intervals.  The two rods are perfectly
rigid and fixed at a specified center-to-center separation, $R$, and
angle $\gamma$.  In all cases, one rod lies parallel to the face
diagonal of the enclosing periodic box \cite{remark1},
while the other
is rotated away from a parallel configuration at
an angle $\gamma$
about the axis connecting the centers of the rods.  In addition, we
introduce mobile multivalent ions of charge +3, and mobile monovalent ions of
charges +1 and -1.  The system is always electrostatically neutral,
with 128 ions of charge +1 to balance the charge on the two rods, one
ion of charge -1 for every additional ion of charge +1 (monovalent or
1:1
salt), and
3 ions of charge -1 for every ion of charge +3 (trivalent or 3:1 salt).

We reference salt concentrations to the total charge on the two
rods (128 electronic charges).
For
example, a \MCconc\ salt concentration of $c_{3:1}=1$ means that the total
charge due
to +3 ions is equal to the total charge on the two rods.
Similarly, a \Saltconc\ salt concentration of $c_{1:1} = 1$ means
that the total charge
due to +1 ions from the monovalent salt is equal to the total
charge of the two rods.

We include two types of pair interactions between particles.  First, we
use the truncated Lennard-Jones
potential to allow for
short-range repulsions.
This introduces the energy scale
$\epsilon$ and the particle size $\sigma$.  Second, we include the
Coulomb interaction, $Z_1Z_2/\varepsilon r_{12}$, where $Z_i$
is the charge on particle $i$ and $\varepsilon$ is the dielectric
constant.
To handle the long-range Coulomb interaction in our system with
periodic boundaries
we use the Particle Mesh Ewald (PME) method
\cite{PETERSEN:95}.

Our simulations are carried out in the
canonical (NVT) ensemble with the temperature fixed at $\kb T = 1.2\epsilon$
using the Langevin thermostat \cite{THIJSSEN:99}.
The monomer number density, $10^{-4} \sigma^{-3}$,
corresponds to a box volume of (109$\sigma)^3$.
The dielectric constant 
is chosen such that
the Bjerrum length (the distance at which the electrostatic
interaction between two electronic charges is equal to the thermal
energy) is $l_{\rm B} \equiv 1/\varepsilon\kb T = 3.2\sigma$.
The separation between charges on the rod (i.e., the monomer separation)
is $l=1.1\sigma$ and the
dimensionless Oosawa-Manning ratio
\cite{OOSAWA:68,MANNING:78} is $l_{\rm B}/l = 2.9$, well above the
threshold for counterion condensation.

The characteristic time scale for the simulation is $\tau=\sqrt{m
\sigma^2/\epsilon}$, where $m$ is the particle mass.  We use the
leapfrog-Verlet integration scheme with a time step of $0.01 \tau$.
Each simulation run is equilibrated (as measured by
the leveling-off of the energy) for at least $10^3\tau$ before we
collect data.  The
force on each monomer is then averaged over
$40\tau$ intervals until we obtain 250--450
average-force data points.
More extended runs were performed for a few systems to
check that our results are not affected by the choice of
equilibration time, time step or data collection interval.
For each simulation run we calculate both the average normal force
per monomer between the two rods
and the average torque on the rods
about the center-to-center axis.  Error bars (indicated by the size of
the points in our figures)
correspond to the
statistical error associated with the average forces at each $40
\tau$ interval.

To obtain the effective interaction potential, or reversible work,
as a function of the
separation $R$ between the two rods, we
should integrate the normal force with distance as the rods are brought from
infinite separation to $R$ at fixed angle $\gamma$.  This is not
possible within our
periodic-boundary-condition simulation.  Instead, we calculate
the reversible work, $\Delta W(R)$,
for bringing the two rods from a fixed reference
separation \cite{remark2}
of 8$\sigma$ to $R$ at fixed $\gamma$.
Similarly, the effective potential as a function of angle, $\Delta
W(\gamma)$, is calculated
by integrating the torque from $90^\circ$ to $\gamma$ at fixed $R$.

\begin{figure}
\centering{
\psfig{file=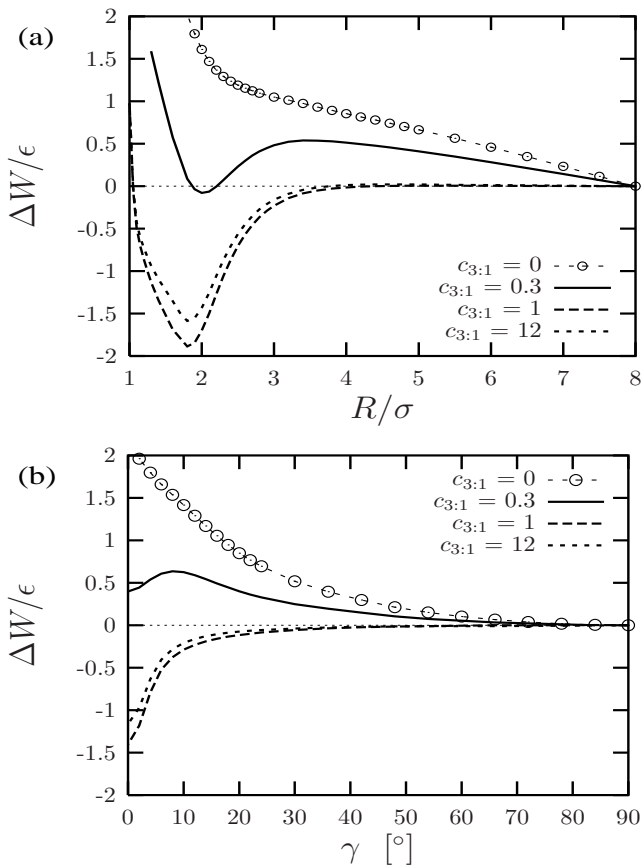,height=4.5in,width=\columnwidth}
}
\caption{
(a) The effective potential, $\Delta W(R)$, between two parallel charged rods
for different concentrations of multivalent salt.  The size of the
solid circles at $c_{3:1} = 0$ corresponds to the error bar for all points
on all curves.
(b) The effective potential, $\Delta W$, as a function of angle,
$\gamma$, for two rods
separated by $R=2.1 \sigma$, for different concentrations of multivalent salt.
}
\label{norm_all}
\end{figure}

We first ask how the effective interaction between two charged rods
depends on the concentration of multivalent salt, in the absence of
monovalent salt.
Fig.~\ref{norm_all}(a) shows a plot of the effective potential per
monomer, $\Delta W$, as a function of separation, $R$, when the two rods are
parallel ($\gamma=0^\circ$).
In the absence of \MCconc\ salt $\Delta W(R)$ is positive for all $R$,
implying that the rods
repel.  When enough \MCconc\ salt is added ($c_{3:1} \approx
0.3$),
the effective
potential develops a global minimum at small $R$; the rods now
attract each other.
This is the threshold \MCconc\ salt concentration for
aggregation; note that this corresponds to approximately only
one-third of the charge on the rods neutralized by trivalent ions.
Beyond this threshold
the attraction increases with multivalent salt
concentration until $c_{3:1} = 1$; this is where the charge
on the rods
is completely neutralized by the trivalent counterions.  Beyond
$c_{3:1} = 1$, the attraction decreases
slightly (short-dashed curve; $c_{3:1} = 12$).
Thus, the magnitude of the attraction is
non-monotonic with \MCconc\ salt \cite{WU.BRATKO.EA:99}.
Similar non-monotonicity is observed in experiments with mixed salts on DNA
solutions \cite{RASPAUD.LIVOLANT:03}.
Though it is imperceptible in the figure, the effective
potential beyond 4.5$\sigma$ is positive for both $c_{3:1} = 1$ and
$c_{3:1} = 12$,
which means that the rods still repel one another at large distance.

The angle dependence of the effective rod-rod interaction is shown in
Fig.~\ref{norm_all}(b) for $R=2.1 \sigma$, near the attractive
minimum.
In the absence of \MCconc\ salt
the rods prefer to be perpendicular since the rods repel one another.
Just above
the threshold for aggregation
(solid curve),
the global minimum is at 90$^\circ$, implying that the preferred aggregated
configuration is a cross.
As the concentration of multivalent salt increases further the
minimum at $\gamma=0^\circ$ deepens, and at $c_{3:1} = 1$ the
rods now prefer to aggregate in a parallel configuration.
When still more multivalent salt is added
the minimum at small angle decreases; this shows again that the
effective potential depends non-monotonically on the concentration of
multivalent salt.

\begin{figure}
\centering{
\psfig{file=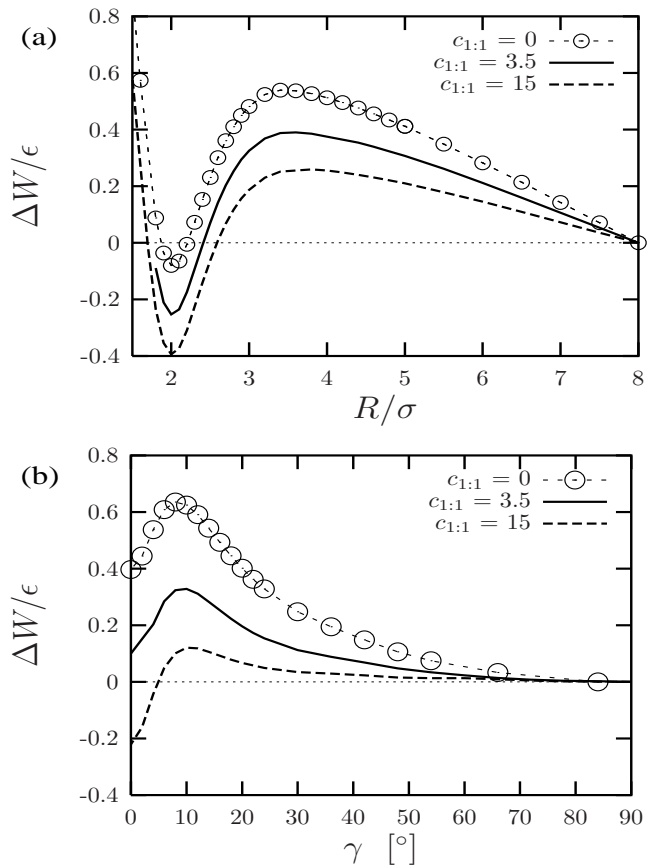,height=4.5in,width=\columnwidth}
}
\caption{
The effective potential between two charged rods, $\Delta W$ as a
function of (a) separation $R$ for parallel rods ($\gamma = 0^{\circ}$)
and (b) angle $\gamma$ for closely-separated rods ($R=2.1 \sigma$),
for different monovalent salt concentrations.  The multivalent salt
concentration is fixed at $c_{3:1} = 0.3$, just above the threshold for
aggregation.  The size of the solid circles at $c_{1:1} = 0$
indicates the error bar for all points on all curves.
}
\label{N13}
\end{figure}

In many-rod systems under conditions for which a pair of rods prefers
to aggregate
perpendicularly, we would expect to find networks, rafts or other structures
where rods cross each other at large angles.  On the other hand, when
a pair prefers to aggregate
in a parallel configuration, we expect to find bundles or
networks of bundles.
Note that for a pair of rods, the preferred angle at which they
aggregate does not vary continuously with increasing
multivalent salt as it would at zero temperature, but rather jumps
from 90$^\circ$ to 0$^\circ$.  This entropic effect
suggests that transitions from network/raft structures to bundles
may be first-order \cite{BORUKHOV.BRUINSMA.EA:03}.

Fig.~\ref{N13} shows the effects of added monovalent salt on
the effective interaction at a multivalent
salt concentration
of $c_{3:1} = 0.3$, just above the threshold for aggregation.
In Fig.~\ref{N13}(a) the attraction
is seen to grow stronger with increasing \Saltconc\ salt.
Furthermore, we find (not shown) that if the
\MCconc\ salt concentration is {\it below} threshold, then \Saltconc\
salt can actually drive the pair to aggregate.
Fig.~\ref{N13}(b) shows that the
metastable minimum at $0^{\circ}$
increases in depth and becomes the global minimum with increasing
\Saltconc\ salt.
This result suggests that adding \Saltconc\ salt to a
solution of charged rods in a network or raft phase can drive the solution
into the bundle phase.  To our knowledge, an experiment has not yet
been designed to test these two effects of added \Saltconc\ salt.

\begin{figure}
\centering{
\psfig{file=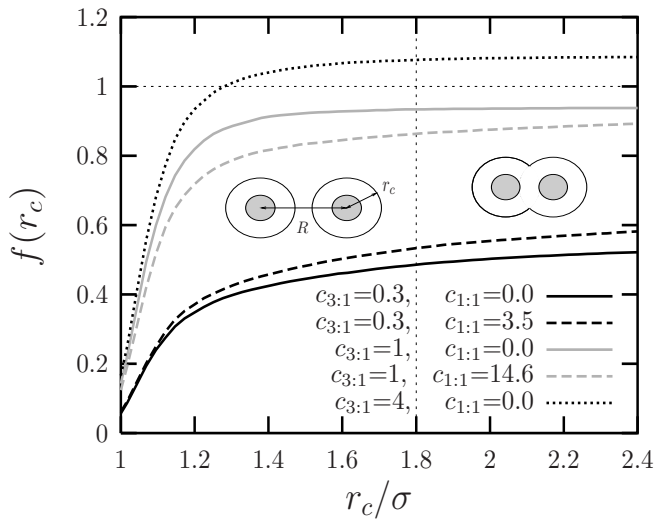,height=2.7in,width=\columnwidth}
}
\caption{
Neutralization fraction $\mathit{f}$ for 2 parallel rods at $R=3.6\sigma$
as a function of the radius of
the encompassing volume $r_{c}$ for several different
concentrations of multivalent and monovalent salts.  The vertical
dotted line corresponds
to $r_{c}=R/2$; to its left, we sum the mobile charges enclosed in two
separate volumes, as sketched in top view, while to the right, we sum
the mobile
charges enclosed in a single volume with a figure-eight cross-section.
}
\label{fr_net}
\end{figure}

To see why monovalent salt enhances the effective attraction, we
measure the neutralization fraction, $\mathit{f}(r_c)$, defined as
the time average of the sum of all mobile charges,
positive and negative, within
radius $r_c$ from either one of the rods, divided by the total charge of
a rod \cite{remark3}.
The sketches in Fig.~\ref{fr_net} depict
a top view of the volume that encompasses the mobile charges included in the
calculation of $\mathit{f}(r_c)$ for a given $r_c$ and separation $R$.
For $ r_{c} <
R/2$ (to the left of the vertical dotted line), there are two separate
enclosing volumes, as sketched, while for $r_{c} > R/2$ (right of
the vertical line), the two enclosing volumes merge into one.
Fig.~\ref{fr_net} shows
$\mathit{f}(r_c)$ for different concentrations
of \Saltconc\ and \MCconc\ salts.  The solid (dashed) curves 
correspond to cases
without (with) \Saltconc\ salt.  Near the \MCconc\ salt threshold
for aggregation,
$c_{3:1} = 0.3$, the solid and dashed black curves show
that $\mathit{f}(r_c)$
increases with added \Saltconc\ salt.  We find that the effective charge on the
rods is reduced by nearby counterions from the \Saltconc\ salt while
the concentration of multivalent counterions near the rods is nearly
unchanged \cite{GONZALEZ-MOZUELOS.OLVERA-DE-LA-CRUZ:03}.  As a
result, the repulsive contribution to the effective interaction
between rods is reduced while the attractive contribution
is unaffected, giving rise to the increase in the net effective
attraction shown in Fig.~\ref{N13}.  (Note that for the black dashed curve,
the screening length--about $6 \sigma$--is
significantly larger than the inter-rod separation of
$R=3.6 \sigma$ and hence not relevant here.)

At high concentrations of multivalent salt ($c_{3:1} \ge 1$),
we find that monovalent salt has the opposite effect--it actually {\it reduces}
the effective attraction.  This is consistent with experiments on DNA solutions
\cite{RASPAUD.LIVOLANT:03}.  This effect also can be understood by
looking at $\mathit{f}(r_c)$.
The gray solid and dashed curves in Fig.~\ref{fr_net} show that
$\mathit{f}(r_c)$ decreases with added \Saltconc\ salt, consistent with earlier
predictions \cite{GONZALEZ-MOZUELOS.OLVERA-DE-LA-CRUZ:03}.
Furthermore, we find
that fewer multivalent ions contribute to $\mathit{f}(r_c)$
upon the addition of \Saltconc\ salt because the multivalent ions stay
in solution in the
form of complexes with monovalent co-ions.  We conclude that co-ions
lure multivalent
ions away from the two rods, causing the attraction to decrease.

The top curve (dotted) in Fig.~\ref{fr_net} shows that
overcharging occurs ($\mathit{f}(r_c)>1$) at sufficiently
high concentrations of multivalent
salt.  Note that $\mathit{f}(r_c)$
crosses 1 at $r_c \approx 1.3\sigma$.  Since $R > 2 r_{c}$ at this point
(the situation sketched to the left of the vertical dotted line),
the total charge enclosed in each cylinder shown is
now positive and the rods are
overcharged \cite{NGUYEN.ROUZINA.EA:00}.
This overcharging saturates with \MCconc\ salt; we find,
for example, that when the multivalent salt concentration is tripled from
$c_{3:1} = 4$ to $c_{3:1} = 12$ there is only a few percent increase
in $\mathit{f}(r_{c})$.
Beyond a certain concentration of \MCconc\
salt, additional multivalent ions stay in solution in the form of
complexes with oppositely-charged monovalent ions.  As a result, we
never observe strong overcharging.

Fig.~\ref{norm_all} shows that overcharging weakens the effective
attraction.  At first glance this is not surprising because
overcharging should increase the Coulomb repulsion
\cite{NGUYEN.ROUZINA.EA:00}.  However, when we increase the \MCconc\ salt
concentration above $c_{3:1} = 1.0$ we find that the normal force between
rods (which we integrate to obtain the effective interaction) remains
negligibly small at large separations, whereas it becomes significantly
less negative at small separations.  This implies that overcharging does
not affect the contribution of the longer-ranged Coulomb repulsion but
appreciably weakens the contribution of counterion-mediated attractions to the
effective interaction.  Fig.~\ref{fr_net}
suggests why the repulsion is not significantly affected by
overcharging: $f(r_{c})$ is approximately as far above unity for
$c_{3:1} = 4$
as it is below unity for $c_{3:1} = 1$.  Thus, the magnitude
of the total charge enclosed within $r_{c}$ is the same in the two
cases although the sign of the charge has flipped.

How then does overcharging weaken the counterion-mediated
attraction?   As the concentration of multivalent salt increases, the
amount of condensed charge increases slightly by accumulating
primarily on the far sides of the rods.
We find that this reorganization of charge (from "bonding" to "antibonding"
regions) leads to a decrease of the attraction.

We thank Markus Deserno and David Reguera for instructive
discussions.  This work was supported by NSF-CHE-0096492 (AJL), 
NSF-CHE-9988651 (WMG), and DE-AC04-94AL85000 (MJS).


\begin{thebibliography}{28}
\bibitem[{rem({\natexlab{a}})}]{remark1}
\bibinfo{note}{Because of periodic boundary conditions our results depend
  slightly on the absolute orientation of the two rods, but none of our
  qualitative conclusions are affected by a particular choice of orientation.}
\bibitem[{rem({\natexlab{b}})}]{remark2}
\bibinfo{note}{At the threshold, \MCconc\ = 0.3, the reversible work for
  bringing the two rods from the box boundary to a separation of $8 \sigma$ is
  at most about 0.5$\epsilon < \kb T$.}
\bibitem[{rem({\natexlab{c}})}]{remark3}
\bibinfo{note}{We have checked that the $\mathit{f}(r_c)$ of each rod is the
  same.}
\end{thebibliography}

\end{document}